\documentclass[twocolumn,showpacs]{revtex4}

\usepackage{epsfig}
\usepackage{dcolumn}
\usepackage{amsmath}

\begin{document}

\title{Do As antisites destroy the ferromagnetism of (Ga,Mn)As?} 

\author{Stefano~Sanvito}
\email{e-mail: ssanvito@mrl.ucsb.edu} 
\author{Nicola~A.~Hill}
\affiliation{\normalsize Materials Department, University of California,
Santa Barbara, CA 93106, USA} 

\date{\today}

\begin{abstract}
The effect of the inclusion of As antisites in the diluted magnetic 
semiconductor (Ga,Mn)As is studied within density functional theory in
the local spin density approximation. In the case of homogeneous
distribution of Mn ions we find that the ferromagnetism is largely 
weakened by the presence of the antisites. This is due to compensation
of the free holes which mediate the long range ferromagnetic order. In
contrast, when two Mn ions are coupled through only one As ion,
ferromagnetic and antiferromagnetic order compete. In this case the magnetic
ground state depends on: i) the position of the As antisites relative to
the Mn,  and ii) the As antisite concentration. We explain our results 
using a model of competing antiferromagnetic super-exchange and ferromagnetic
double-exchange via localized Zener carriers. The strong dependence of the 
ferromagnetic order on the microscopic configuration accounts for the large 
variation in experimental data.
\end{abstract}

\maketitle

In the past ten years the study of diluted magnetic semiconductors (DMS) 
has been strongly revitalized since the discovery of ferromagnetic 
order in In$_{1-x}$Mn$_x$As \cite{Ohno1} and Ga$_{1-x}$Mn$_x$As 
\cite{Ohno2,Ohno3,Ohno4}. The potential utility of room temperature 
ferromagnetism in a semiconductor based system is enormous. On one hand,
DMS could be replace the existing metallic magnetic 
elements in storage media \cite{Prinz}. 
Here their better compatibility with existing semiconductors and MBE-growth technology 
over their metallic counterparts is a great advantage. On the other hand
DMS can be used as spin injectors into semiconductors \cite{Aws1,Sch1}.
This is a critical step in the physical realization of quantum
computation based on the spin degree of freedom in a solid state device 
\cite{DDV}. 
In fact, although it has been shown that spin can be coherently transported over 
several micrometers in GaAs \cite{Aws2,Aws3}, its injection from metallic 
magnetic contacts is largely unsuccessful if not impossible \cite{Schm1}. 
This is due to the large mismatch between the resistances of the metallic 
contacts and the semiconductor. The use of DMS solves this problem \cite{Aws1,Sch1}
and (Ga,Mn)As together with GaAs and (Al,Ga)As represent to date the most
promising material system for the injection, storage and manipulation of
spins in semiconductors.

Although there is general agreement on the carrier- (hole-) mediated origin 
of the ferromagnetism in (Ga,Mn)As, the detailed mechanism is still debated. The Zener
model in the mean field approximation \cite{Diet1} provides a good starting
point. However its prediction of the Curie temperature as a function of Mn
and hole concentration largely overestimates the actual value if a dynamic 
description of the Mn spins is considered \cite{McD}. Moreover recent density
functional theory (DFT) results \cite{us1} show that the $p$-$d$ coupling in 
(Ga,Mn)As is very strong and the mean field approximation cannot be completely 
justified.
Finally DFT calculations based on the coherent-potential approximation (CPA) 
\cite{Akai} suggest that the ferromagnetic order in (Ga,Mn)As may result from the 
competition between ferromagnetic double-exchange and antiferromagnetic 
super-exchange interactions.

Note that all these models agree on the following important 
points: i) Mn$^{2+}$ ions
substitute the Ga$^{3+}$ cations in the zincblende lattice providing local
$S=5/2$ spins, and ii) there are free holes in the system although the actual
concentration is much smaller than the density of Mn ions \cite{Ohno3,Ohno4}. 
This latter point suggests that some mechanism of compensation must be
taking place, most likely the presence of intrinsic donor defects such as As 
antisites (As$_\mathrm{Ga}$) or Ga vacancy-interstitial As pairs 
(V$_\mathrm{Ga}$-As$_i$), which are usually found in low-temperature 
GaAs \cite{Dab}.
In the above models the compensation mechanism is implicitly 
incorporated in the mean field description of the GaAs valence band, since the 
hole concentration is a free parameter. As a result the local effects of 
magnetic and chemical disorder are neglected. 

In this paper we consider explicitly the effects of the inclusion of
As$_\mathrm{Ga}$ in (Ga,Mn)As at different dilutions, and study how the chemical
environment modifies the magnetic interaction between the Mn ions. We perform
DFT calculations within the local spin density approximation (LSDA), and
consider large GaAs cells in which Mn ions and As$_\mathrm{Ga}$ are inserted. 
We study the dependence of the magnetic coupling between Mn ions on: i) Mn
concentration, ii) As$_\mathrm{Ga}$ concentration, iii) relative positions
of the Mn ions and the As$_\mathrm{Ga}$ defects. The main result of our
analysis is that, for a uniform distribution of Mn ions, As$_\mathrm{Ga}$ 
antisites strongly weaken the ferromagnetic coupling. However, at least for 
moderate concentrations, the compensation does not follow the expected nominal 
valences of Mn and As (that is one As$_\mathrm{Ga}$ does not completely compensate 
the holes from two Mn ions). 
Moreover when two Mn ions are separated by only one As atom
(so that they occupy two corners of a zincblende tetrahedron with an As atom in the 
center), ferromagnetic coupling is possible even far above compensation
if As antisites are located at the other two tetrahedral positions to form
Mn$_2$As$_3$ complex (see Fig.\ref{FIG3}c).

Our calculations are performed using the code {\sc siesta} \cite{Siesta1,Siesta2,Siesta3}, 
which is an
efficient implementation of DFT-LSDA based on pseudopotentials and a numerical 
localized atomic orbital basis set. This method combines good accuracy and 
small computational cost compared to other methods based on plane-waves.
We have described in a previous paper \cite{us1} the calculation details and
the prescriptions one needs to optimize the basis set and the pseudopotentials.
Here we just mention that we use Troullier-Martins pseudopotentials 
\cite{TM} with non linear core corrections \cite{Lou82} and Kleinman-Bylander
factorization \cite{KB1}, and the Ceperley-Alder \cite{CA} form of the 
exchange-correlation potential.

We construct 64 (cubic) and 32 (rectangular) atom GaAs cells in which we 
include two Mn ions and a variable number of As$_\mathrm{Ga}$ antisites. 
These correspond to Mn concentrations of $x$=0.0625 (to date 
the largest concentration obtained experimentally) and $x$=0.125 respectively. 
We investigate two
different Mn configurations for each concentration: 
1) the Mn ions occupy positions as far apart as possible 
(i.e. the corner and the middle of the cubic cell), 
2) the Mn atoms occupy two corners of a tetrahedron and are coordinated
through a single As ion (see Fig.\ref{FIG3}). We call these two configurations 
{\it separated} and {\it close} respectively. 
The cells are then periodically repeated using 18 $k$-points in the 
irreducible cubic Brillouin zone. Note that the periodic boundary conditions fix
the magnetic coupling between Mn ions that occupy equivalent positions in adjacent 
cells to be ferromagnetic but that the coupling between the two Mn ions in the
same cell can be ferromagnetic or antiferromagnetic. We calculate the ground-state 
properties of these systems for both ferromagnetic (FM) and antiferromagnetic 
(AF) alignment of the Mn ions within the cells. 

\begin{figure}[ht]
\centerline{\epsfig{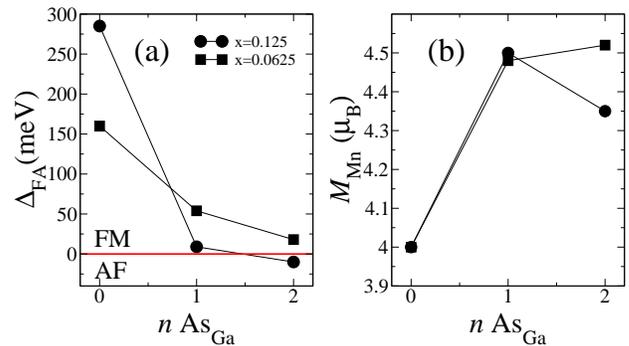}}
\caption{(a) Energy difference between AF and FM alignments, $\Delta_\mathrm{FA}$, 
and (b) magnetization per Mn ion, $M_\mathrm{Mn}$, as a function of the number of 
As$_\mathrm{Ga}$ antisites in the cell: {\it separated} configuration. 
The horizontal line denotes the division between FM and AF alignment.}
\label{FIG1}
\end{figure}

The energy differences between the AF and FM aligned configurations, $\Delta_\mathrm{FA}$,
and the magnetization per Mn ion, $M_\mathrm{Mn}$, for the {\it separated} arrangement
are presented in Fig.\ref{FIG1} as a function of the number of As$_\mathrm{Ga}$ 
antisites. The magnetization per Mn ion is defined to be half of the magnetization of the 
cell, calculated in the FM aligned phase.

First we note that the ferromagnetic coupling is strongly weakened by
As$_\mathrm{Ga}$ antisite doping. This is consistent with the picture of
magnetic coupling mediated by free carriers (holes): As$_\mathrm{Ga}$ antisites
contribute electrons into the system and therefore compensate the holes. 
We also note that in the case of no antisites the
ferromagnetic order is much stronger in the case of large Mn concentration,
which again supports the model of carrier-induced ferromagnetism.
Within this model, the ferromagnetic coupling should disappear when
the compensation is complete. However the figure suggests
that the compensation mechanism does not follow the nominal atomic valence, 
since a single As$_\mathrm{Ga}$ antisite per cell is not sufficient to destroy the 
ferromagnetic coupling. 
Above compensation (one As$_\mathrm{Ga}$ for two Mn ions) antiferromagnetic 
coupling is obtained for large Mn concentration, while the system stays 
ferromagnetic at low concentration. This is consistent with the onset of 
antiferromagnetic super-exchange coupling, the mechanism which is 
believed to be 
responsible for the magnetic order in the II-VI DMS \cite{26}, at compensation. 
Super-exchange
is a short range interaction and therefore is less important in the low
concentration limit where the Mn ions are well separated. 
It is worth noting that for $x$=0.125 and large antisite
concentrations ($n$~As$_\mathrm{Ga}$=2) the magnetic order is quite sensitive 
to the actual position of the antisite respect to the Mn ions. 
Different As$_\mathrm{Ga}$ arrangements may result either in ferromagnetic or
antiferromagnetic coupling.

\begin{figure}[ht]
\centerline{\epsfig{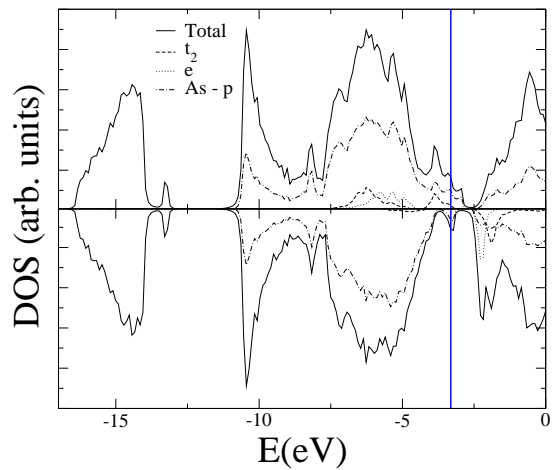}}
\caption{DOS for (Ga,Mn)As obtained for a 64 atom unit cell with two Mn ions and
one As$_\mathrm{Ga}$ antisite: {\it separated} configuration and ferromagnetic
alignment. The upper graph represents the majority and the lower the minority
spins.}
\label{FIG2}
\end{figure}
If we now consider the magnetization per Mn ion (Fig.\ref{FIG1}b) we note that 
there is a monotonic increase of the total magnetic moment on antisite doping. This can
be understood by looking at the density of states of Fig.\ref{FIG2} (see also
Ref.\onlinecite{us1}).
At the Fermi energy the DOS of the majority spin band is the result of the
hybridization between the Mn $d$ states with $t_2$ symmetry and the $p$ states
coming mainly from As$_\mathrm{Ga}$. In fact the two donor levels of isolated 
As$_\mathrm{Ga}$ in GaAs are known to lie 0.54~eV and 0.75~eV above the GaAs valence 
band \cite{Dab} and to overlap with the 
exchange split Mn $d$ levels of low dilution (Ga,Mn)As \cite{us1}.
In contrast the DOS of the minority spin band is dominated by antisite
levels, the first Mn $d$ states lying far above $E_\mathrm{F}$. Therefore by
increasing the antisite concentration, $E_\mathrm{F}$ is shifted towards higher
energy, thus increasing the occupation of the Mn $d$ shell in the majority
spin band. This enhances the total magnetic moment of the system. Finally note 
that a magnetization of 4.5$\mu_\mathrm{B}$ is in very good agreement with 
recent x-ray magnetic circular dichroism measurements \cite{Ohl1}.

Let us now turn our attention to the {\it close} configuration. In this case we
expect the short range super-exchange to play a more important r\^ole since the 
Mn ions are much closer to each other. Moreover the magnetic coupling within a cell is
expected to be quite sensitive to the position of the antisites with respect to the
Mn ions. We consider the three different situations sketched in Fig.\ref{FIG3}:
a) the antisites are far from the Ga$_2$Mn$_2$As$_1$ complex, b) one antisite occupies
a tetrahedral site (Ga$_1$Mn$_2$As$_2$), c) two antisites occupy the 
tetrahedral sites (Mn$_2$As$_3$).
\begin{figure}[ht]
\centerline{\epsfig{file=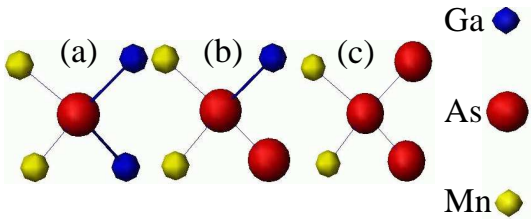,scale=0.3,angle=0}}
\caption{{\it Close} Mn configuration. The three antisite arrangements
considered in the text: a) Ga$_2$Mn$_2$As$_1$, b) Ga$_1$Mn$_2$As$_2$,
c) Mn$_2$As$_3$.}
\label{FIG3}
\end{figure}
In Fig.\ref{FIG4} we show $\Delta_\mathrm{FA}$ and $M_\mathrm{Mn}$ as a
function of the As$_\mathrm{Ga}$ concentration for the different antisite
arrangements of
Fig.\ref{FIG3}. For the sake of brevity we present data only for $x$=0.0625, noting
that similar conclusions can be made also for $x$=0.125.
\begin{figure}[ht]
\centerline{\epsfig{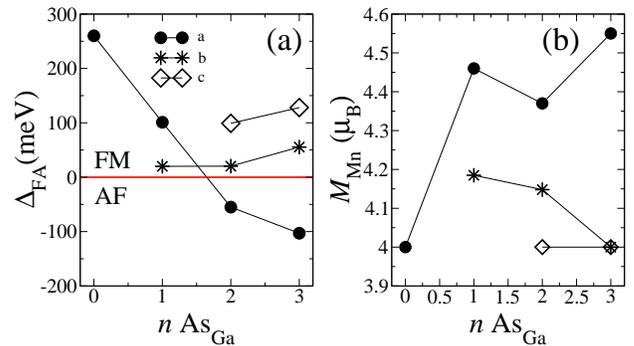}}
\caption{(a) Energy difference between the AF and FM alignments 
$\Delta_\mathrm{FA}$, 
and (b) magnetization per Mn ion $M_\mathrm{Mn}$ as a function of the number of 
As$_\mathrm{Ga}$ antisites in the cell: {\it Close} Mn arrangement. The symbols
$\bullet$, $ * $ and $\diamond$ represent arrangements (a), (b) 
and (c) of Fig.\ref{FIG3} respectively.}
\label{FIG4}
\end{figure}
From Fig.\ref{FIG4} it is very clear that, when the As$_\mathrm{Ga}$ defects are in the
vicinity of the Mn ions (configurations (b) and (c) of Fig.\ref{FIG3}), the system 
is almost insensitive to the As$_\mathrm{Ga}$ concentration. In contrast when 
the As$_\mathrm{Ga}$ antisites are far from the Mn ions the magnetic coupling
undergoes a FM to AF transition with increasing antisite concentration. 

Consider first configuration (a). In this case both the $\Delta_\mathrm{FA}$ 
and $M_\mathrm{Mn}$ curves look very similar to the curves we found for the 
{\it separated} arrangement (Fig.\ref{FIG1}). However, in stark contrast with
the {\it separated} case, an AF alignment can now also be found for {\it large}
As$_\mathrm{Ga}$ concentrations. Recalling the fact that the Mn $d$ shell is
more than 
half-filled, and that it is antiferromagnetically coupled with the As 
$p$ shell of the intermediate atom \cite{us1}, we propose that super-exchange 
coupling stabilizes the AF phase as soon as the free-holes are completely 
compensated. However this mechanism is extremely short range and therefore effective 
only when the Mn ions are separated by one As.

We now turn our attention to situations (b) and (c). In both cases the
ferromagnetic alignment is stable and almost insensitive to the total 
As$_\mathrm{Ga}$ concentration. This is a strong indicator that the dominant 
interaction in these cases is one that depends only on the local chemical properties. To
shed some light on this aspect we present the results of M\"ulliken population analyses 
\cite{Mul,us1} of the system.
\begin{table}[hbtp]
\begin{tabular}{cccccccc}
\hline
$n$~As$_\mathrm{Ga}$ & Type & Mn-$d_\uparrow$ & Mn-$d_\downarrow$ & As-$p_\uparrow$ &
As-$p_\downarrow$ & As-$p$ & As \\ \hline \hline
2 & a  & 4.74 & 0.70 & 1.55 & 1.63 & 3.18 & 4.92 \\ 
2 & b  & 4.74 & 0.71 & 1.50 & 1.67 & 3.17 & 4.95 \\ 
2 & c  & 4.72 & 0.75 & 1.44 & 1.71 & 3.15 & 4.97 \\ \hline \hline

3 & a  & 4.76 & 0.68 & 1.57 & 1.63 & 3.20 & 4.93 \\ 
3 & b  & 4.74 & 0.74 & 1.55 & 1.65 & 3.20 & 4.97 \\ 
3 & c  & 4.73 & 0.75 & 1.49 & 1.69 & 3.17 & 4.99 \\ \hline \hline
\end{tabular}
\caption{M\"ulliken atomic and orbital populations for the Mn ions and the
intermediate As atom of the complexes of Fig.\ref{FIG3}. The Mn concentration 
is $x$=0.0625. 
The symbols $\uparrow$ and $\downarrow$ correspond to majority and minority 
spin respectively. The populations are in units of the electronic charge $|e|$.}
\label{TAB1}
\end{table}
The M\"ulliken orbital (or atomic) population is the projection of the charge 
density onto a particular orbital (or atom). Although the absolute values depend on
the chosen basis set, it is a useful quantity for understanding the charge
distribution within a given material. In table \ref{TAB1} we present the 
M\"ulliken orbital population for the two Mn ions and the intermediate As ion of
the complexes of Fig.\ref{FIG3}. The most obvious feature is the larger
polarization of the As-$p$ orbitals in Mn$_2$As$_3$ compared with 
Ga$_2$Mn$_2$As$_1$.  This is accompanied by a small decrease in Mn-$d$ polarization,
although this latter effect could be an artifact from the overlap component of the
orbital population \cite{us1}. It is also interesting to note that the total
atomic population of the As atom increases going from (a) to (b) to (c), while 
the $p$ component of the population decreases.
Therefore if we  start from the situation in which two As$_\mathrm{Ga}$
antisites are located far from the Mn-As-Mn complex, then we move each antisite
in turn to one of the two other corners of the tetrahedron, then i) the
charge on the middle As atom increases, ii) the spin-polarization of the $p$
shell of the middle As atom increases, and iii) the total population of the $p$-shell
of the middle As atom decreases. And most importantly the magnetic coupling changes
from antiferromagnetic to ferromagnetic.

In 1960 de~Gennes observed \cite{DG} that in an antiferromagnetic crystal
the presence of a bound carrier (electron or hole) which is Zener coupled to 
the local spins {\it always} induces a distortion in the antiferromagnetic
lattice. In that spirit we propose that the observed transition from
antiferromagnetic to ferromagnetic coupling between Ga$_2$Mn$_2$As$_1$
and Mn$_2$As$_3$ results from the onset of ferromagnetic double-exchange
coupling mediated by a bound Zener carrier.
This is further supported by recent calculations \cite{us2} in which we show
that the actual ground state for Ga$_1$Mn$_2$As$_2$ is not one of perfect
ferromagnetic alignment, but instead the Mn magnetic moments are canted with respect to
each other.

In conclusion we have shown that the As$_\mathrm{Ga}$ antisites'
inclusion in (Ga,Mn)As can result in a variety of different behaviors depending
on the microscopic arrangement of the Mn ions and the As$_\mathrm{Ga}$ antisites.
In particular we have shown that when the Mn ions are uniformly distributed in
the crystal, As$_\mathrm{Ga}$ antisites weaken the ferromagnetic order. 
In contrast if the Mn ions occupy two corners of the zincblende tetrahedron
several magnetic arrangements are possible depending on the positions of the
antisites. In particular ferromagnetic coupling is obtained if antisites
occupy the other positions in the tetrahedron. This suggests that a way to
obtain high $T_\mathrm{c}$ in (Ga,Mn)As is either to reduce the As antisite
concentration, or to deposit Mn and As antisites with a strongly 
inhomogeneous distribution. 

This work made use of MRL Central Facilities supported by the National Science 
Foundation under award No. DMR96-32716. This work is supported by the DARPA/ONR 
under the grant N0014-99-1-1096, by ONR grant N00014-00-10557, by NSF-DMR under 
the grant 9973076 and by ACS PRF under the grant 33851-G5.

\end{document}